\documentclass[preprint,5p,times,twocolumn]{elsarticle}

\usepackage{amssymb}
\usepackage{lineno}
\journal{Nuclear Physics B}

\begin{document}

\begin{frontmatter}



\title{The Two-Screen Measurement Setup to Indirectly Measure Proton Beam Self-Modulation in AWAKE}

\author[CERN,TUG]{M. Turner}

\author[CERN,CZK]{B. Biskup}

\author[CERN]{S. Burger}

\author[CERN]{E. Gschwendtner}

\author[CERN]{S. Mazzoni}

\author[CERN]{A. Petrenko}

\address[CERN]{CERN, Geneva, Switzerland}

\address[TUG]{Graz University of Technology, Graz, Austria}

\address[CZK]{Czech Technical University, Prague, Czech Republic}

\begin{abstract}
The goal of the first phase of the AWAKE \cite{AWAKE1,AWAKE2} experiment at CERN is to measure the self-modulation \cite{SMI} of the $\sigma_z = 12\,\rm{cm}$ long SPS proton bunch into microbunches after traversing $10\,\rm{m}$ of plasma with a plasma density of $n_{pe}=7\times10^{14}\,\rm{electrons/cm}^3$. The two screen measurement setup \cite{Turner2016} is a proton beam diagnostic that can indirectly prove the successful development of the self-modulation of the proton beam by imaging protons that got defocused by the transverse plasma wakefields after passing through the plasma, at two locations downstream the end of the plasma.
This article describes the design and realization of the two screen measurement setup integrated in the AWAKE experiment. We discuss the performance and background response of the system based on measurements performed with an unmodulated Gaussian SPS proton bunch during the AWAKE beam commissioning in September and October 2016. We show that the system is fully commissioned and adapted to eventually image the full profile of a self-modulated SPS proton bunch in a single shot measurement during the first phase of the AWAKE experiment.
\end{abstract}


\end{frontmatter}


\section{Introduction}
\subsection{The AWAKE experiment}
The Advanced Proton-Driven Plasma Wakefield Acceleration Experiment (AWAKE) \cite{AWAKE1,AWAKE2} is a proof-of-principle R\&D experiment at CERN that uses a $400\,\rm{GeV/c}$ proton bunch, with $3\times10^{11}\,\rm{protons/bunch}$, from the CERN SPS to create GV/m plasma wakefields over meter scale distances. To excite strong plasma wakefields efficiently, the proton beam bunch length has to be in the order of the plasma wavelength $\lambda_{pe}$. In the AWAKE experiment we use a $10\,\rm{m}$ long rubidium vapor source with a density of $n_{pe}=7\times10^{14}\,\rm{electrons/cm}^3$ (which corresponds to a plasma wavelength of $\lambda_{pe}=1.2\,\rm{mm}$) and the SPS proton bunch has a length of $\sigma_z = 12\,\rm{cm}$. Hence, the experiment relies on the development of the self-modulation instability \cite{SMI} (SMI) to modulate the long proton bunch into micro-bunches spaced at the plasma wavelength. The SMI is seeded by the sharp proton beam edge (or sudden turn-on of the plasma) created by overlapping the proton bunch with a short ($100\,\rm{fs}$) laser pulse  ionizing the rubidium. As the SMI develops, transverse plasma wakefields periodically focus and defocus the proton beam creating a micro-bunch structure.  
\begin{figure}[htb!]
\centering
		\includegraphics[width = \columnwidth]{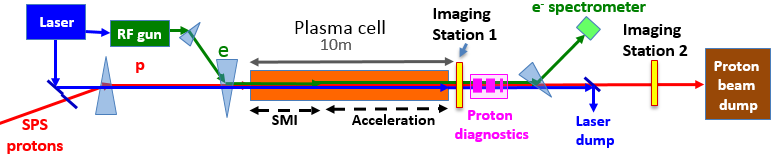}
		\caption{Schematic layout of the AWAKE experiment}
		\label{fig:layoutAWAKE}
\end{figure}
The goal of the first phase of the AWAKE experiment (starting in December 2016) is to prove that the SMI developed successfully and that GV/m plasma wakefields were created. 

In the second phase of the experiment (starting in late 2017), $10-20\,\rm{MeV}$ electrons will be injected into the wakefield, and accelerated to an energy of several $\rm{GeV}$. A schematic layout of the AWAKE experiment is shown in Figure \ref{fig:layoutAWAKE}.

In this paper we describe in detail the design choices of the measurement setup designed to indirectly prove proton beam self-modulation in the AWAKE experiment at CERN. We discuss the performance and background response of the system based on measurements taken during the AWAKE proton beam commissioning in 2016 with an unmodulated Gaussian proton beam and no plasma.

\subsection{Challenges and Requirements of the Two-Screen Measurement System}
\label{sec:measurement}

\begin{figure}[htb!]
\centering
		\includegraphics[width = \columnwidth]{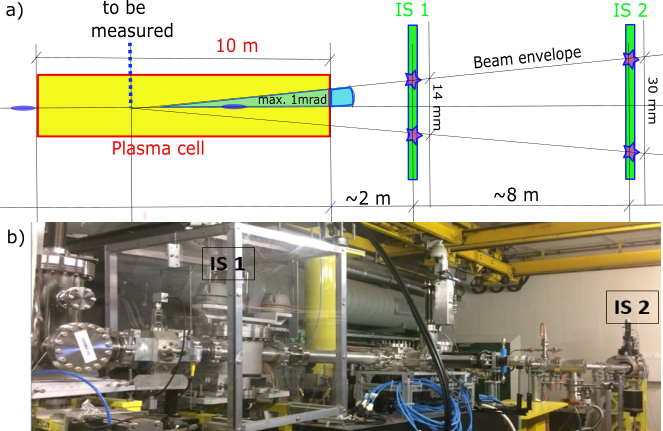}
		\caption{a) Schematic drawing of the two-screen setup. b) An image of the AWAKE experimental area, showing the two installed imaging stations.}
		\label{fig:layout}
\end{figure}
The idea of the two screen measurement \cite{Turner2016} is to image the strongest defocused protons (defocused by the plasma wakefield) approximately $2\,\rm{m}$ and $10\,\rm{m}$ downstream the end of the plasma (see Figure \ref{fig:layout}a) using two imaging stations (IS). From this measurement, we can calculate the maximum proton defocusing angle. Measuring maximum defocusing angles in the order of $1\,\rm{mrad}$ (instead of $0.05\,\rm{mrad}$  without plasma) indirectly proves that strong plasma wakefields were present in plasma. Additionally, by using the position and angle of the maximum defocused protons, we determine at which position along the plasma these protons got their radial kick.

\begin{figure}[htb!]
\centering
		\includegraphics[width = \columnwidth]{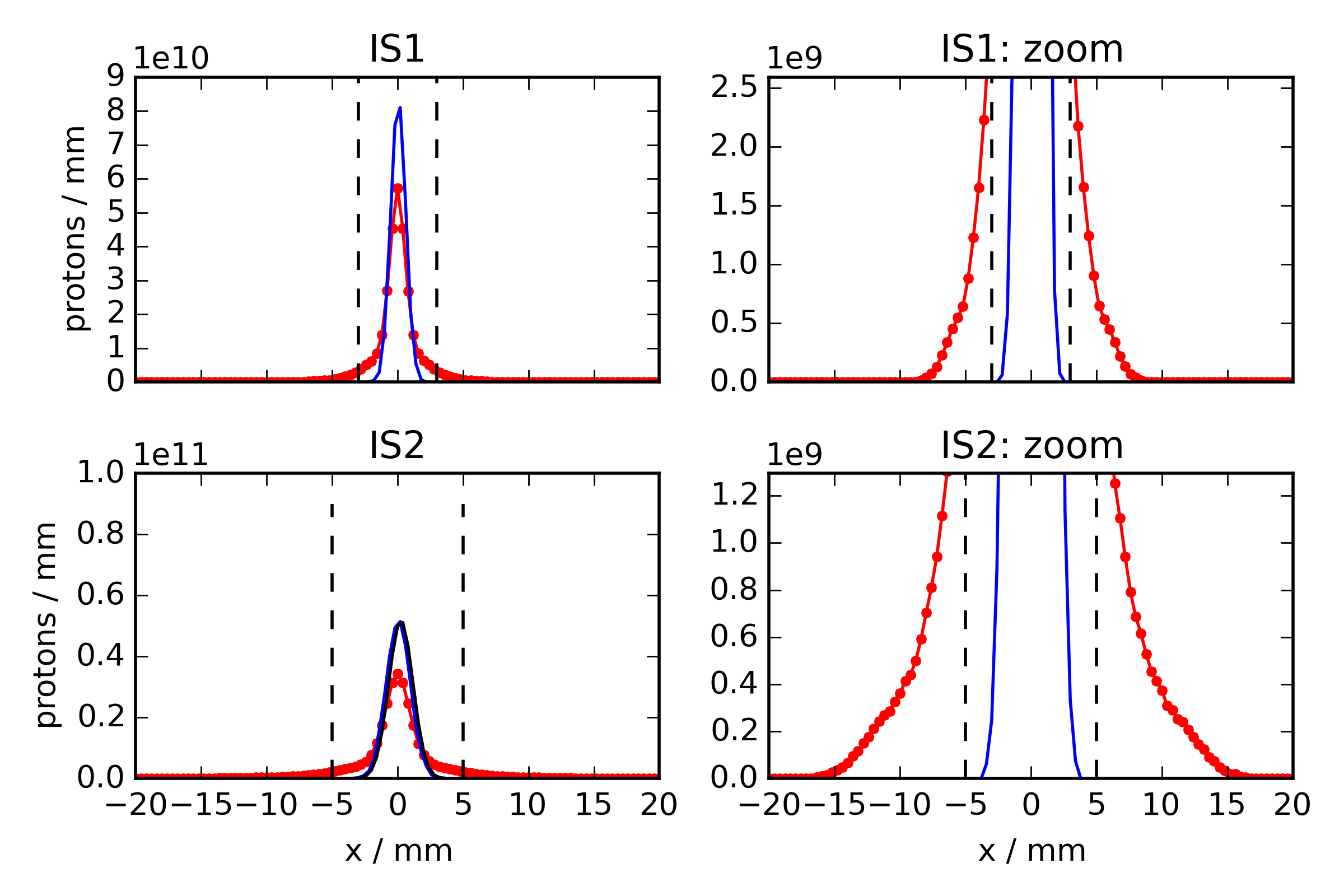}
		\caption{Expected proton beam distribution with (red lines with dots) and without plasma (blue lines) on the first and second imaging station. The 2D image is integrated over the vertical plane. Vertical dashed lines show the size of the holes that were cut into the imaging screens.}
		\label{fig:beamdistr}
\end{figure}

From plasma simulations, performed with 2D3v quasistatic LCODE \cite{LCODE1,LCODE2}, and using the AWAKE baseline parameters \cite{AWAKE1} we expect a self-modulated transverse proton beam distribution like the one shown with red dots in Figure \ref{fig:beamdistr}. The blue line in Figure \ref{fig:beamdistr} shows the unmodulated Gaussian proton beam distribution that we expect to measure on the screens if there is no plasma present. Since we want to determine the maximum defocusing angle of the self-modulated proton bunch, it is crucial that the system can measure the outermost - or strongest defocused- protons \cite{TurnerNapac2016}, because the outermost protons experienced the highest product of wakefield strength times interaction distance.

The challenges of measuring a proton beam distribution like the one shown with red dots in Figure \ref{fig:beamdistr} are:
\begin{itemize}
\setlength\itemsep{0em}
 \item We want to image the outermost edge of the defocused beam but the beam core is 4 orders of magnitude more intense.
 \item We must be able to measure the self-modulated proton bea edge as well as the unmodulated proton beam (in case of no plasma) with the same system.
 \item The setup works in a radiation environment ($\approx10^{12}\,$ high energy hadrons per cm$^2$ per year).
\end{itemize}
Based on this requirements we developed two specialized beam imaging stations (IS), based on available systems at CERN. Both IS include a choice of screens (including screens consisting of more than one material and screens with a hole), that emit optical light either via Optical Transition Radiation (OTR) or Scintillation and the setup uses a radiation-resistant analogue camera to capture the emitted light.
Figure \ref{fig:layout}b shows the AWAKE experimental area, including the two imaging stations (IS) for the two screen measurement.\\

Further requirements on these detector stations are as follows:
\begin{itemize}
 \item The screens withstand the impact of the $400\,\rm{GeV/c}$ proton beam with $3\times10^{11}\,\rm{protons}$ every $30\,\rm{seconds}$.
 \item The optical light emitted by the proton beam edge (expected from plasma simulations: $\approx10^6\,\rm{protons/mm}^2$ ) gives a clear signal on the camera.
 \item  We require a measurement resolution of $0.4\,\rm{mm}$ to measure the maximum defocusing angle within an uncertainty of 10\% and the origin of the defocused particles with a precision of $\pm1.3\,\rm{m}$.
\end{itemize}

\section{The beam imaging station in AWAKE}
\begin{figure}[htb!]
\centering
		\includegraphics[width = 0.5\columnwidth]{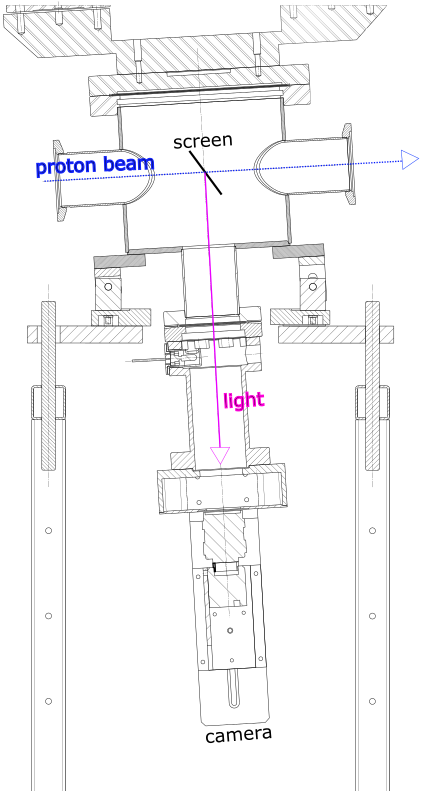}
		\caption{Technical drawing of the first proton beam imaging station.}
		\label{fig:BTV}
\end{figure}

Two imaging stations (IS1 and IS2) measure the transverse proton beam distribution $2\,\rm{m}$  and $10\,\rm{m}$ downstream the end of the plasma. The imaging stations, as shown in Figure \ref{fig:BTV}, consist of the support, vacuum vessel, screen insertion device, light emitting screens, illumination system, filter wheel, lens and camera. The light emitted by the imaging screens goes through a filter and a focusing lens (IS1: focal length = $50\,\rm{mm}$; IS2: focal length = $35\,\rm{mm}$) before it reaches the CCD image sensor. 
The analog interlaced video feed (camera: WATEC 902-H3 ULTIMATE (CCIR)) is then digitized to $400 \times 300\,\rm{pixels}$ \cite{ISCARD}.   
The field of view is 26.98 x 34.9 mm on IS1 and 57.51 x 71.42 mm on IS2.  
The camera is not synchronized to the proton bunch extraction from the CERN SPS. Independent of the proton bunch, the camera triggers every $20\,\rm{ms}$. This does not affect the amount of light captured from Optical Transition Radiation (OTR) since OTR is emitted instantaneous. But it affects the amount of light captured from a scintillating screen because the light is emitted over tens of milliseconds. Previous measurements \cite{HIRADMAT}, performed with an unsynchronized camera, showed that the camera captured significantly less light (up to -50\%) in approximately 30\% of the images captured with a $1\,\rm{mm}$ thick Chromox screen. Consequently, we expect to reject approximately 30\% of the performed measurements.

\begin{enumerate}
\setlength\itemsep{0em}
\item The first imaging station holds two different screens (screen size = $6\times6\,\rm{cm}$): 
\begin{itemize}
\setlength\itemsep{0em}
\item SiAg: $0.3\,\rm{mm}$ thick Silicon coated with Silver (OTR).
\item Chromox: $1\,\rm{mm}$ thick $\rm{Al}_2\rm{O}_3:\rm{CrO}_2$ with a hole $r = 3\,\rm{mm}$ (Scintillator).
\end{itemize}
\item The second imaging station holds three different screens (screen size = $10\times10\,\rm{cm}$):
\begin{itemize}
\setlength\itemsep{0em}
\item Chromox: $1\,\rm{mm}$ thick (Scintillator).
\item Chromox: $1\,\rm{mm}$ thick with a hole $r = 5\,\rm{mm}$ (Scintillator).
\item Combined Screen: $1\,\rm{mm}$ thick Chromox with an $r = 5\,\rm{mm}$, $1\,\rm{mm}$ thick Aluminum insertion at the center (Scintillator + OTR).
\end{itemize}
\end{enumerate}
The size of the holes was chosen based on Figure \ref{fig:beamdistr} to cut off most of the beam core.
Screens and filters can be changed remotely. The filters have an opening of $1\,\rm{inch}$ and are limiting the aperture in the optical line from the screens to the camera for the first imaging station. In the second imaging station the acceptance is limited by the $1\,\rm{inch}$ lens. Filters in the filter wheels were chosen to be transmitting 1\% and 0.1\% of the captured light for the first and 1\%, 0.1\% and 0.01\% for the second imaging station.

\subsection{Selected screen material properties}
As discussed in section \ref{sec:measurement}, the imaging screens must fulfill the following requirements:
\begin{itemize} 
\item All screens withstand the impact of the proton beam.

We tested 15 different screen materials in the HiRadMat \cite{HIRADMATFACILITY} facility at CERN \cite{HIRADMAT}. All screens withstood the proton beam impact ($2 \times 10^{11}$ protons/bunch at $440\,\rm{GeV/c}$), but a change of color was noticed on two undoped Alumina samples. Based on the results of these measurements we choose to measure the defocused proton beam edge with a $1\,\rm{mm}$ thick Chromox screen ($\rm{Al}_2\rm{O}_3:\rm{CrO}_2$) because it created the second strongest signal on the camera and did not blur the signal as much as the tested $3\,\rm{mm}$ Chromox screen.

\begin{figure}[htb!]
\centering
		\includegraphics[width = \columnwidth]{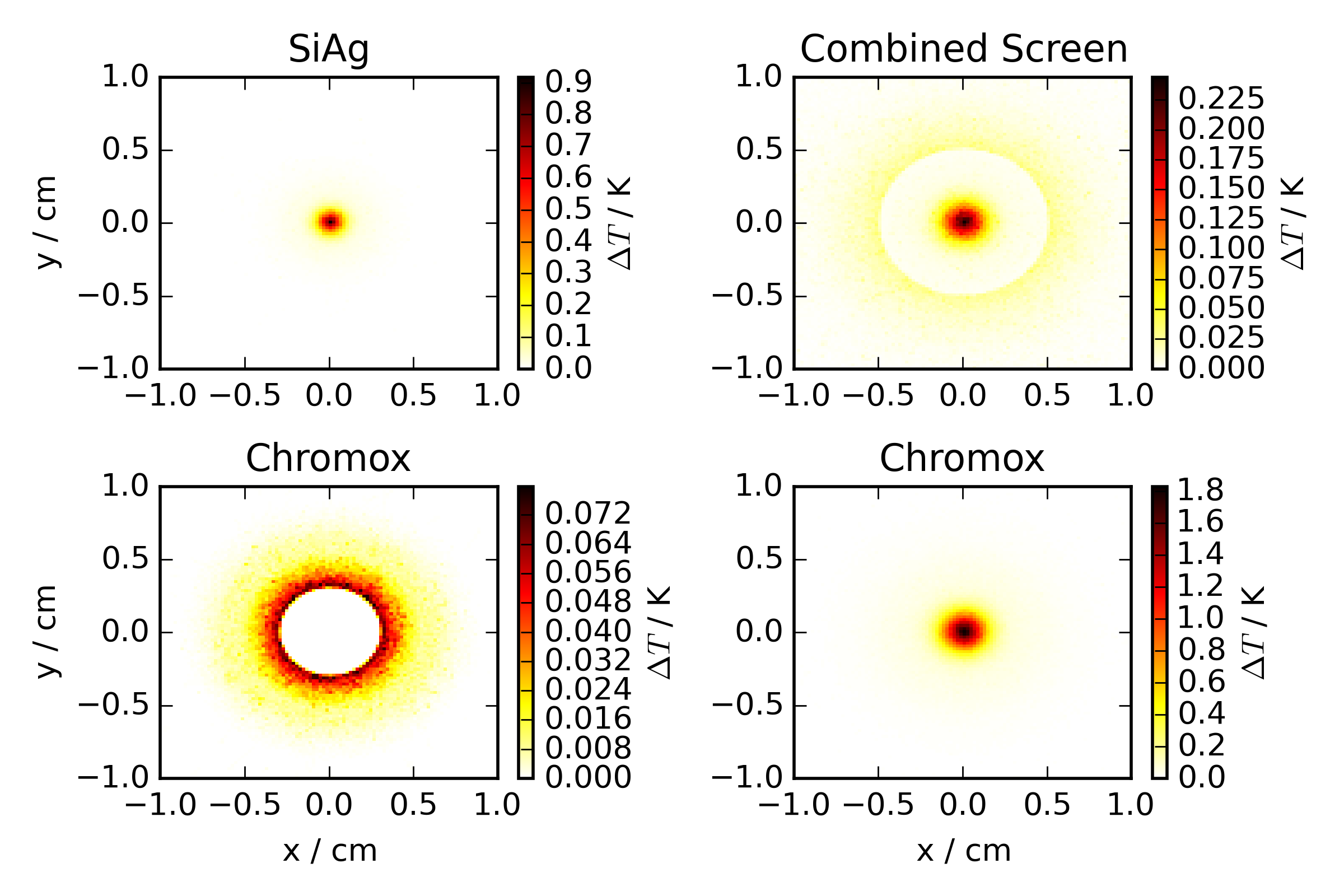}
		\caption{Temperature rise in the imaging screens. The images on the left show the screens of the first imaging station and on the right of the second.}
		\label{fig:trise}
\end{figure}
As the self-modulated proton beam will pass through the imaging screens, it will deposit energy in the screen material. We calculated the temperature rise in the imaging screens, based on the energy deposit of the proton beam simulated with FLUKA \cite{FLUKA}. In the case of Chromox we used the density $\rho$ and heat capacity $c$ of Aluminum ($\rho_{\rm{Alu}} = 2.7\,\rm{g/cm}^3$, $c_{\rm{Alu}} = 0.897\,\rm{J/(kg K)}$). We can neglect the Silver coating of the Silver coated Silicon screen, because the coating layer is only a few nm thick. ($\rho_{\rm{Si}} = 2.3\,\rm{g/cm}^3$, $c_{\rm{Si}} = 0.703\,\rm{J/(kg K)}$).

Figure \ref{fig:trise} shows that the maximum temperature rise in the screens ranges from $0.07$ to $1.8\,\rm{K}$ per self-modulated proton beam pulse. The AWAKE experiment will get beam every $30\,\rm{seconds}$ and the screens will handle this small, locally limited temperature increase, as the melting point of the screens ranges from 660 to 2000 degrees Celsius.

\item The light emitted by the proton beam edge (expected from plasma simulations: $\approx 10^6 \rm{protons/mm}^2$ ) gives a clear signal on the camera.

FLUKA calculations show that a $400\,\rm{GeV/c}$ proton looses about $1\,\rm{MeV}$ when traversing $1\,\rm{mm}$ of Chromox. From simulations we expect $\sim10^{6}\,\rm{protons/mm}^2$ in the defocused proton beam edge and that $1\,\rm{MeV}$ of deposited energy emits $\sim10^4$ photons \cite{LY} over $4\pi$. The acceptance of our imaging system is limited by the $1\,\rm{inch}$ opening of the lens and the distance between the screen and the lens ($\approx 50\,\rm{cm}$). We estimate that using the lowest demagnification approximately $3000$ photons arrive on one pixel imaging $\approx90\times90\,\mu \rm{m}^2$ (we know from experience with similar setups that there should be a minimum of 1000-2000 photons/pixel to get a clear signal).

\item Screen resolution better than $0.4\,\rm{mm}$.

The screen resolution is estimated by twice of the size that one pixel images and ranges from $180\,\mu\rm{m}$ on IS1 to $400\,\mu\rm{m}$ on IS2.
 
\end{itemize}

\section{Optimized imaging screens}
The self-modulated proton beam density projected onto a 2D screen extends over 4 orders of magnitude (see Figure \ref{fig:beamdistr}). The limited dynamic range ($2-3$ orders of magnitude) of the CCD camera makes the detection of the defocused proton beam edge challenging. One solution to measure the defocused edge is to cut a hole into the imaging screen (see vertical dashed lines in Figure \ref{fig:beamdistr}), so that the dense beam core can pass without interacting with the screen. Following this idea two $1\,\rm{mm}$ Chromox screens with a hole radius of $3\,\rm{mm}$ and $5\,\rm{mm}$ were installed in two imaging stations in the AWAKE tunnel.\\
\begin{figure}[htb!]
\centering
		\includegraphics[width = 0.4\columnwidth]{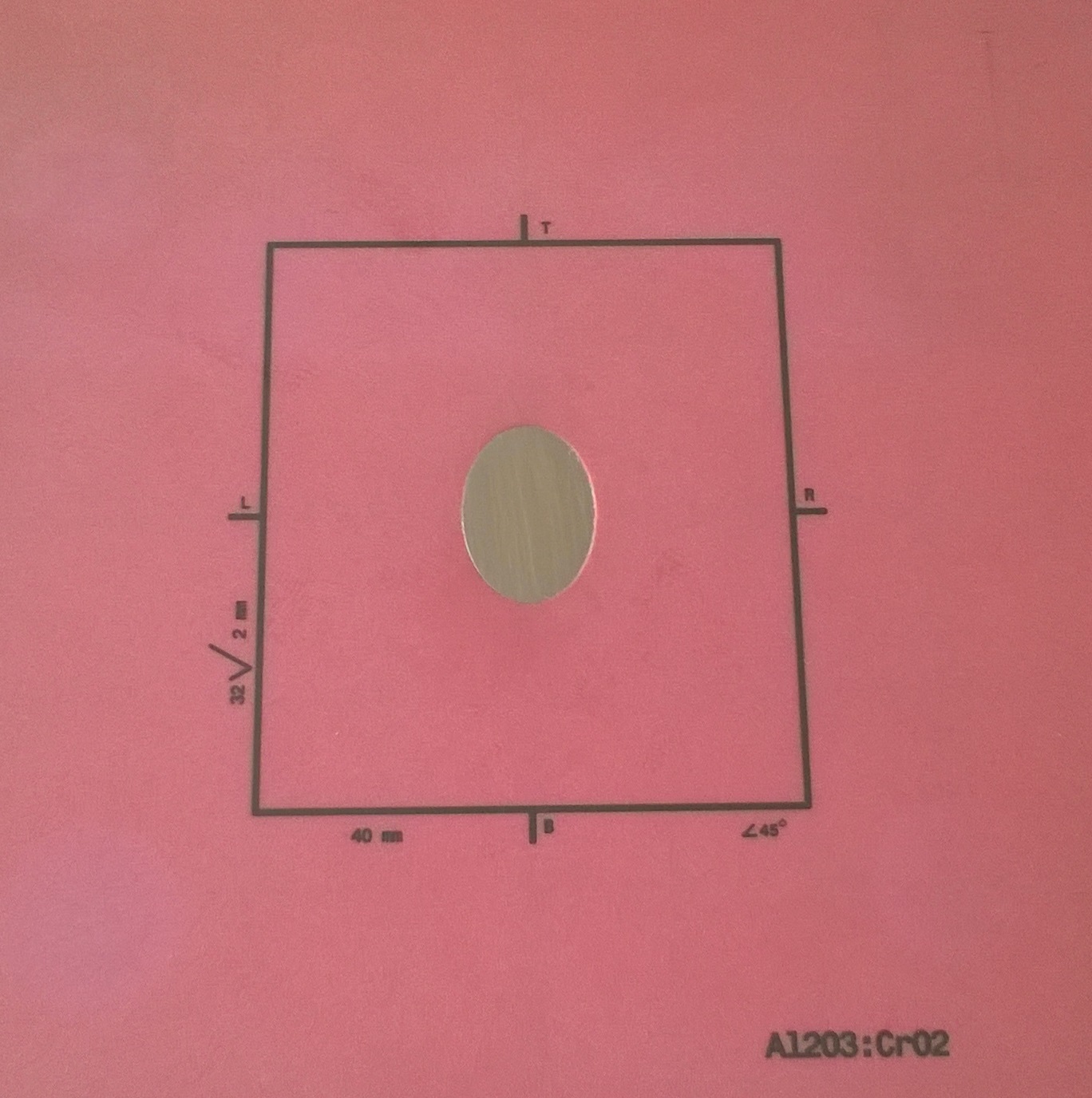}
		\caption{Beam imaging screen combining the high-light-yield scintillating material Chromox for measuring the defocused proton beam edge together with the low-light-yield OTR emitting material Aluminum for the measurement of the intense beam core.}
		\label{fig:combscreen}
\end{figure}
\begin{figure}[htb!]
\centering
		\includegraphics[width = 1\columnwidth]{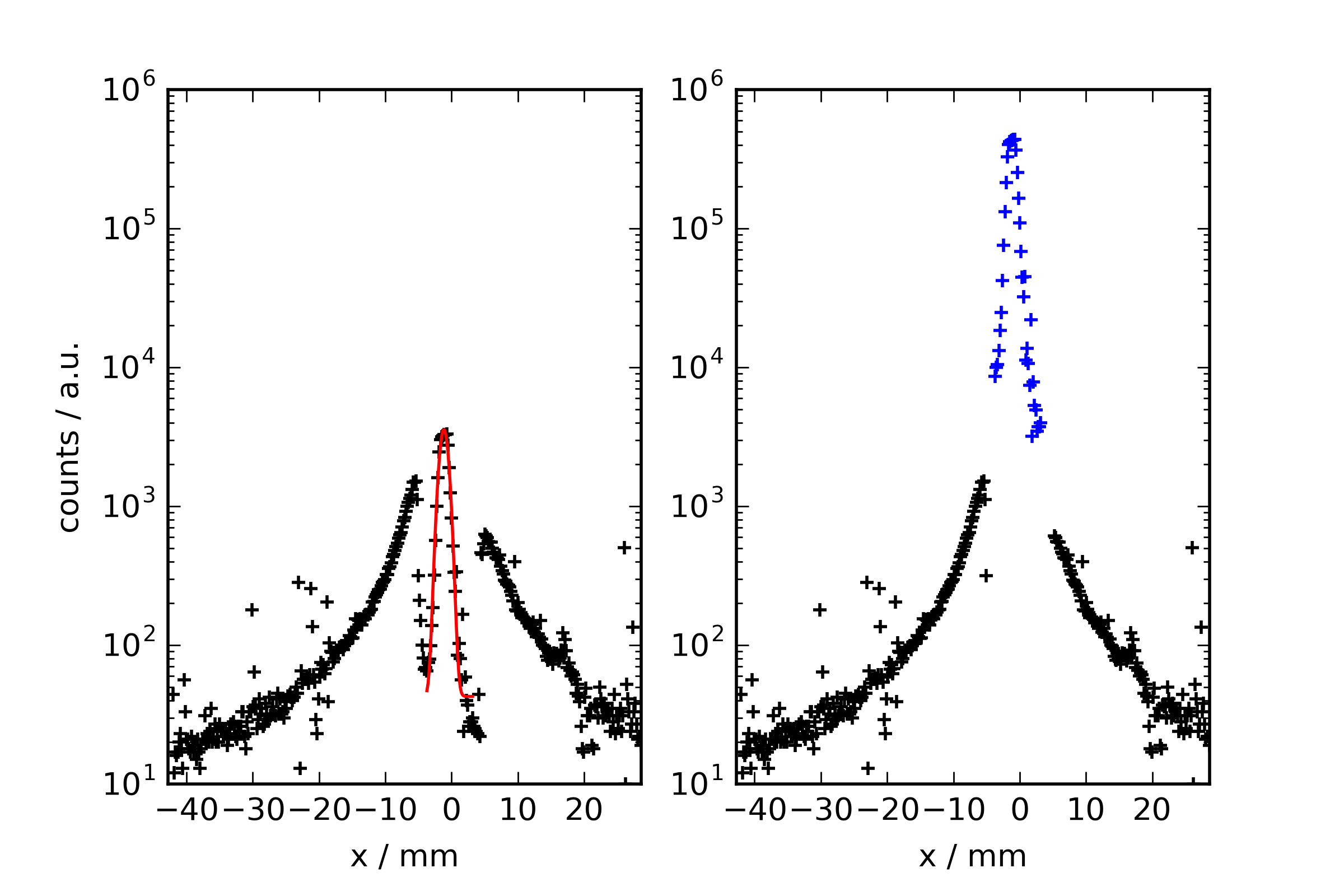}
		\caption{The left image shows the central pixelline of a measurement with the combined screen of IS2. The right image shows the same data, but the beam core is boosted by the light emission ratio of Chromox over Aluminum. The red line of the left image shows a Gaussian fit of the measured beam core.}
		\label{fig:combscreen_m}
\end{figure}

\subsection{Combined Screen}
Another possibility of handling the intense beam core is a combined screen (see Figure \ref{fig:combscreen}). 
In the combined screen, the hole at the center of the scintillating Chromox screen was filled with the OTR emitting material Aluminum. A combined screen was installed in the third slot of the IS2. The left side of Figure \ref{fig:combscreen_m} shows a measurement of the combined screen during AWAKE proton beam commissioning in October 2016. There was no plasma, so we imaged an unmodulated Gaussian proton beam from the CERN SPS with a momentum of $400\,\rm{GeV/c}$ and ${1\,\times\,10^{11}\,\rm{protons}}$  per bunch.

During the beam commissioning, we captured light emitted by the Aluminum and Chromox screens on our cameras. After normalizing to the extracted proton beam intensity and the used optical filter we calculated that with our setup the Chromox screen emits $2700\pm400$ times more photons than the Aluminum screen.

The expected transverse proton beam sigma $\sigma_r$ at the second imaging station (see blue line in Figure \ref{fig:beamdistr}) is $\sigma_r = 0.92\,\rm{mm}$. We measure the radial beam sigma in Figure \ref{fig:combscreen_m} by fitting a Gaussian distribution to the beam core and obtain from the fit $\sigma_r = 0.7\pm 0.02\,\rm{mm}$ which agrees within the resolution of the measurement. For large radial positions (r $> 5 \,\sigma_r$) proton bunches are not perfectly Gaussian so we did not perform a fit on the data of the surrounding Chromox.

On the right side of Figure \ref{fig:combscreen_m} we boost the light emitted by the Aluminum at the center of the screen by 2700 (ratio between the light emission of Aluminum and Chromox) to reconstruct the full beam profile. Figure \ref{fig:combscreen_m} shows that the combined screen can measure beam intensities of almost five orders of magnitude in a singe-shot measurement using a standard CCD camera.

The defocusing angle of the defocused protons and the saturation point of the self-modulation instability can only be analysed after we performed measurements with plasma.

\subsection{Beam and Screen Alignment}
As illustrated in Figure \ref{fig:beamdistr}, the proton beam should be aligned to the center of the hole in the screen within $0.5\,\rm{mm}$ (expected proton beam jitter: $\sigma_r = 100\,\mu\rm{m}$ ). We aligned the imaging screens by mechanically moving the IS tanks, after the final proton beam trajectory was set.

\section{Signal to background measurements}
 The light emission of scintillators is proportional to the energy deposited in the material. Using FLUKA, we simulated the energy deposition of the self-modulated proton beam in the $1\,\rm{mm}$ thick Chromox screen of the second imaging station.
 To understand if inserting an imaging screen on IS1 influences the energy deposit on a screen inserted in IS2 (i.e. because of secondary particles production by the first screen), we studied the energy deposit in the Chromox screen of IS2 for three different scenarious: 
\begin{enumerate}
\setlength\itemsep{0em}
 \item IS1: Screen out.
 \item IS1: SiAg in.
 \item IS1: Chromox in.
\end{enumerate}
Simulations did not show any measureable increase of energy deposit for scenario 2) and 3).

\begin{figure}[htb!]
\centering
		\includegraphics[width = \columnwidth]{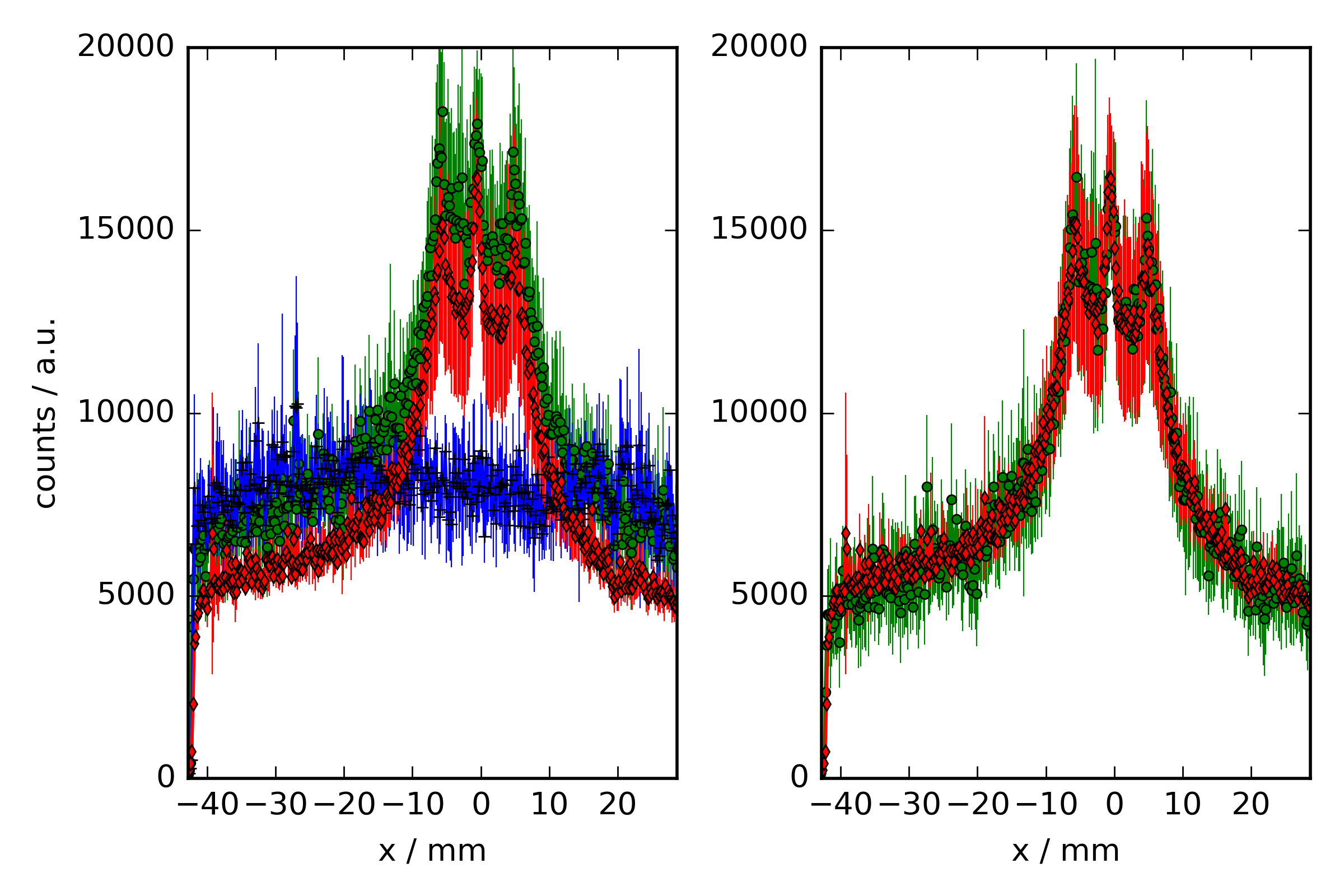}
		\caption{Proton beam measurements of the combined screen performed during AWAKE beam commissioning. The dots of every curve represents the average of 8 measurements and the errorbars show the standard deviation. The data is integrated in the vertical plane. Red diamonds show measurement on the combined screen on IS2 for when no screen was inserted at IS1, the green 'o' when the SiAg screen was used in IS1 and the black '+' show the meaurement on IS2 (no screen inserted) while the SiAg screen was used in IS1.}
		\label{fig:bg}
\end{figure}

During proton beam commissioning of the AWAKE experiment in September 2016, we benchmarked those simulation results. We imaged a unmodulated Gaussian proton beam with $1\times 10^{11}$ protons, using the combined screen of the second imaging station (IS2). The results are presented in Figure \ref{fig:bg}. The red diamonds in the left part of the figure present the measurement for when no screen was inserted at IS1 (scenario 1), and the green 'o' when the SiAg screen was used in IS1 (scenario 2). In contrast of the simulations, we notice that the background increased by $\approx1800\,\rm{counts/bin}$. 

As a consequence we studied the origin of the background by capturing empty images on IS2 (no screen inserted), while the SiAg screen was inserted in IS1 (black '+') and we measured the same level of background as in scenario 2). Additionally the measured background level was independent of the filter used, so we concluded that secondary particles produced by the first imaging screens directly impact on the CCD chip of the camera.

The nuclear interaction length of Silicon is $46.52\,\rm{cm}$ \cite{PDG}. The thickness of the first imaging screen that we inserted was $0.3\,\rm{mm}$ which means that 1 out of 2500 protons has a nuclear interaction with the screen material and can create secondary particles. Combining this estimate with a previous one on Coulomb scattering of the protons \cite{Turner2016}, we do not expect any measurable change of the proton beam profile and we also confirmed this with measurement shown of the right image of Figure \ref{fig:bg}.

This Figure \ref{fig:bg} shows the overlap of the measurement for when no screen was inserted at IS1 (scenario 1, red diamonds) and when the SiAg screen was used in IS1 (scenario 2, green 'o'). We subtracted the pedestal of $1800\,\rm{counts}$ from this measurement and achieved an overlap in the height of the signal. In the right image of Figure \ref{fig:bg} we observe neither a blurring nor a measurable distortion of the proton beam profile.

It was not possible to study the effect of the Chromox screen in IS1 on the measurement with IS2, because the unmodulated proton beam that we used during commissioning entirely passes through the $r = 3\,\rm{mm}$ hole of the Chromox screen in IS1.

\section{Conclusions}
We have validated the design of two beam imaging stations installed downstream the plasma in the AWAKE experiment. These stations are designed to measure the defocused proton beam edge of the self-modulated proton beam by using a $1\,\rm{mm}$ thick Chromox screen. We chose the scintillator Chromox because it withstands the impact of the $400\,\rm{GeV/c}$ SPS proton bunch and emits enough light to detect a proton beam density of $10^6\,\rm{protons/mm}^2$. To measure the self-modulated proton beam edge next to the four orders of magnitudes more intense beam core, we developed a combined screen which consists of an Aluminum screen surrounded by Chromox.
Imaging an unmodulated Gaussian proton beam on the combined screen showed that the imaging system can measure proton beam intensities of almost five orders of magnitudes with a standard CCD camera.

The background signals of the imaging stations are understood and were measured in the AWAKE experiment environment: secondary particles produced by the first screen can impact on the camera of the second imaging station and increase the background level by $\approx1800\,\rm{counts/bin}$ (in a vertically integrated beam image). Inserting a screen in the first imaging station does not worsen the measured proton beam profile on the second station.

The measurement setup is fully commissioned and ready for the measurement of the self-modulation instability of the proton beam in the AWAKE experiment.

\appendix
\section{References}



\end{document}